\documentclass[a4paper]{article}
\usepackage{amsmath,amssymb,amsthm,amssymb,graphicx}
\usepackage[T1]{fontenc}
\usepackage[latin1]{inputenc}
\title{Quantum correlations and dynamics from classical random fields valued in complex Hilbert spaces}
\author{Andrei 
\\
School of Mathematics and Systems Engineering\\
University of V\"axj\"o, S-35195, Sweden}

\begin{document}

\maketitle
\abstract{One of the crucial differences between mathematical
models of classical and quantum mechanics is the use of the tensor
product of the state spaces of subsystems as the state space of
the corresponding composite system. (To describe an ensemble of
classical composite systems one uses random variables taking
values in the Cartesian product of the state spaces of
subsystems.) We show that, nevertheless, it is possible to
establish a natural correspondence between the classical and
quantum probabilistic descriptions of composite systems. Quantum
averages for composite systems (including entangled) can be
represented as averages with respect to classical random fields.
It is essentially what Albert Einstein was dreamed of. Quantum
mechanics is represented as classical statistical mechanics with
infinite-dimensional phase space. While the mathematical
construction is completely rigorous, its physical interpretation
is a complicated problem (which will not be discussed in this
paper).}

\section{Introduction}

Nowadays it is commonly accepted that the use of the {\it tensor
product of the state spaces} of subsystems as the state space of
the corresponding composite system is one of the main
distinguishing features of QM. It is especially important in
quantum information theory where systems in entangled states play
an fundamental role.\footnote{While it is sufficiently well
studied mathematically, entanglement is still quite mysterious
physically. Its widely used interpretation as the evidence of
``nonlocal correlations'' can not be considered as completely
satisfactory, see, e.g., \cite{AKK}- \cite{AKK6} and especially \cite{KKK}-\cite{IK} for recent
debates. Therefore clarification of the structure  of its
mathematical description may have important consequences. This
paper is a step in this direction.} In this paper  we do not
discuss extremely complicated problems related to interpretations
of quantum mechanics. We proceed in the framework of  mathematical physics.

There are known two models \cite{IK} for computations of averages
for ensembles of composite systems: a) classical probability model
(due to Kolmogorov \cite{K}) based on integrals, b) quantum
probability model (due to von Neumann \cite{VN}, see, e.g.,
\cite{GR}, \cite{Ozawa}, \cite{CK} for the modern treatment of the
problem) based on traces of self-adjoint operators. We show that,
in spite of a rather common opinion, quantum correlations for
observables on subsystems of a composite systems can be
represented  as correlations with respect to a classical
(Gaussian) random fields. Moreover, dynamics of quantum
correlations induced by Schr\"odinger's equation can be reduced to
dynamics of correlations for a classical ``prequantum'' stochastic
processes.

For non-composite systems theory  has been developed in a series
of author's papers \cite{EKHR}- \cite{KH6}. It is known under the
name {\it Prequantum Classical Statistical Field Theory} (PCSFT).
In this theory ensembles of quantum particles are represented by
classical random fields, probability distributions on a Hilbert
space. We remark that appealing to classical random fields is
rather common in various attempts to create a kind of classical
statistical mechanics which reproduces predictions of QM. We can
mention stochastic electrodynamics, e.g., \cite{R2}, \cite{BOY},
or the semiclassical model, e.g., \cite{C1}-- \cite{C3}. Bohmian
mechanics also contains a kind of classical field, the pilot wave.
However, in this model randomness is coupled to particles and not
to fields. The same can be said about Nelson's stochastic QM
\cite{N} and its generalization due to Mark Davidson \cite{MD},
\cite{MD1} as well as the recent prequantum model of `t Hooft
\cite{H1}, \cite{H2}, see also Thomas Elze \cite{EL1}. Physically
the present paper belongs to the domain of  ``quantum mechanics as
emergent phenomenon'', cf. mentioned papers of `t Hooft and Elze
as well as models which were recently created Rusov et al. \cite{Rus} and 
Kisil \cite{Kis}.

However, as was already pointed,  we prefer not to go in the
debate on a physical meaning of the proposed mathematical
construction. Our aim was to unify two mathematical descriptions
of averages, classical and quantum. This aim was approached via
representation  of quantum correlations by Gaussian integrals over
the Hilbert space $H_1\times H_2,$  where $H_i.i=1,2,$ are the
(Hilbert) state spaces of the subsystems. On the other hand, we
could not totally escape the interpretation problem. The main
message from our mathematical construction is that a quantum pure
state of a composite system should be considered not as a ``state
vector'' belonging to the tensor product $H_1 \otimes H_2,$ but as
non-diagonal block of an operator acting in the Cartesian product
$H_1 \times H_2$ see (\ref{HHUURR}).
 This operator, say $D,$ is the {\it covariance
operator} of the prequantum classical random field. We remind that
consideration of a {\it density operator as the  covariance
operator of the corresponding prequantum random process} is the
crucial point of PCSFT, see \cite{EKHR}- \cite{KH6}. In this paper
we extend this approach to composite systems.

We point out to another approach providing a possibility to represent quantum averages
by operating only with classical probability distributions, namely, quantum tomographic 
approach, see Manko et al. \cite{M1}- \cite{M3}.

Results of these paper were shortly announced in \cite{EPL}, where they were presented 
on the physical level of rigorousness. Unfortunately,  such a formal presentation induced 
(to be totally  honest) a mistake -- unfortunately, in the basic equality (\ref{01}) coupling the 
quantum correlation with the nontrivial term of the classical field correlation. In this paper we 
proceed in the rigorous mathematical framework, the basic formula and its consequences were 
corrected. Formulas of the present theory 
are more complicated than of one given in \cite{EPL}, but they are mathematically correct.

\section{Statistical models of classical and quantum mechanics}

Everywhere in this papers Hilbert spaces are separable. Let $H_j,
j=1,2, $ be (real or complex) Hilbert spaces. We denote the space
of bounded linear operators from $H_1$ to $H_2$  by the symbol
${\cal L}(H_1, H_2).$  Let $H$ be a (real or complex) Hilbert
space. We denote the space of self-adjoint bounded operators in
$H$ by the symbol ${\cal L}_s(H).$

\subsection{Classical model}

$\; \; \;$ a).  States are represented by points of some set $M$
(state space).

b). Physical variables are represented by functions $f: M \to
{\bf R}$ belonging to some functional space
$V(M).$\footnote{The choice of a concrete functional space
$V(M)$ depends on various physical and mathematical factors. In classical mechanics for systems with the finite
number of degrees of freedom $M$ is chosen as the phase space ${\bf R}^{2n}; V(M)$ is the space of smooth functions.}

c). Statistical states are represented by probability measures on
$M$ belonging to some class $S(M).$

d). The average of a physical variable (which is represented by a function $f \in V(M))$ with respect to a statistical state (which is
represented by a probability measure  $p \in S(M))$ is given by
\begin{equation}
\label{AV0} < f >_p \equiv \int_M f(\phi) d p(\phi) .
\end{equation}
By using the language of probability
theory we can say that there is given a random vector
$\phi(\omega),$ where $\omega$ is a random parameter, taking
values in $M.$ Then $\langle f\rangle_\phi = E f(\phi(\omega))=
\langle f\rangle_p.$ Here and everywhere below $E$ denotes
classical mathematical expectation (average).

If the state space $M$ is a space of functions, e.g., $M=L_2({\bf
R}^3),$ then $M$-valued  random vectors are called  {\it random
fields.} For each $\omega,$ $\phi(\omega)$ is a function of $x \in
{\bf R}^3: \phi(x, \omega).$

e). If systems
$S_i, i=1,2,...,k,$ have state spaces $M_i,$  respectively, then
the composite system $S=(S_i)_{i=1}^k$ has state space
$M=M_1\times\ldots \times M_k,$ the Cartesian product of the state spaces
$M_i.$ Ensembles of $S$-systems are described by random vectors in
$M: \phi(\omega)=(\phi_1(\omega), \ldots, \phi_k(\omega))$ (or equivalently by probability measures on $M.)$  A
trivial, but important, remark is that in general  components of
$\phi(\omega)$ are not independent. There are nontrivial
correlations between them. The best way to describe these
correlations is to use the {\it covariance operator}  (it will be
defined little bit later).

\medskip

A {\it classical statistical model} is a pair $M=(S(M),
V(M)).$
\subsection{Quantum case}

Let $H$ be a complex Hilbert space.

\medskip

a). States (pure) are represented by classes of normalized vectors
of $H$ with respect to the equivalence relation: $\psi_1= e^{i
\theta}  \psi_2.$

a). Physical observables are represented by operators
$\widehat{A}: H \to H$ of  the class ${\cal L}_{{\rm s}} (H).$ (To
simplify considerations, we shall  consider only quantum
observables represented by bounded operators.)

b). Statistical states are represented by density operators. The class of such operators
is denoted by  ${\cal D}(H).$

d). Average of a physical observable (which is represented by
the operator $\widehat{A} \in {\cal L}_{{\rm s}} (H))$ with respect to a
statistical state (which is represented  by the density operator $\rho
\in {\cal D} (H))$ is given by von Neumann's formula
\begin{equation}
\label{AV1}
<A >_\rho \equiv \rm{Tr}\; \rho \widehat{A}
\end{equation}

e). If quantum systems $S_i, i=1,2,...,k,$ have the state spaces
$H_i,$  respectively, then the system $S=(S_i)_{i=1}^k$ has the
state space $H_1\otimes\ldots \otimes H_k,$  the tensor product of
state spaces $H_i.$

\medskip

The {\it quantum statistical model} is the pair $N_{\rm{quant}}
=({\cal D}(H), {\cal L}_{s}(H)).$

\medskip

At the first sight the gap between classical statistical mechanics
and quantum mechanics is huge \cite{VN}. Impossibility to reduce
quantum averages to classical averages is the main source of the
great ideological difference between classical and quantum
probabilistic descriptions.

\section{Gaussian measures on real and complex Hilbert spaces}

\subsection{Real case}

Let $W$ be a real Hilbert space. Let $A \in {\cal L}_s(W).$

We start with derivation of the basic mathematical formula which
was used in \cite{EKHR}- \cite{KH6}. We will calculate the
Gaussian integral of the quadratic form
\begin{equation}
\label{QF}f_A (\phi)= (A\phi, \phi).
\end{equation}
Consider  a $\sigma$-additive Gaussian measure $p$ on the
$\sigma$-field of Borel subsets of $W.$ This measure is determined
by its covariance operator $B:  W \to W$ and mean value $m \in W.$
For example, $B$ and $m$ determine the Fourier transform of $p$
$$
\tilde p(y)= \int_W e^{i(y, \phi)} dp (\phi)=
e^{\frac{1}{2}(By, y) + i(m, y)}, y \in W.
$$
In what follows we restrict our considerations to {\it Gaussian
measures with zero mean value}: $ (m,y) = \int_W(y, \psi) d p
(\psi)= 0 $ for any $y \in W.$ Sometimes there will be used the
symbol $p_B$ to denote the Gaussian measure with the covariance
operator $B$ and $m=0.$ We recall that the covariance operator $B$
is defined by its bilinear form
\begin{equation}
\label{CO} (By_1, y_2)=\int (y_1, \phi) (y_2, \phi) dp(\phi),
y_1, y_2 \in W,
\end{equation}
and  it has the following properties: a) $B \geq 0,$ i.e., ($By,
y) \geq 0, y \in W;$ b) $B$ is a self-adjoint operator, $B \in
{\cal L}_s(W);$ c) $B$ is a trace-class operator and ${\rm Tr}\;
B=\int_W ||\phi||^2 dp(\phi).$ It is {\it dispersion}  of the
probability $p.$ Thus for Gaussian probability we have
$\sigma^2(p)= {\rm Tr}\; B.$ We remark that the list of properties
of the covariation operator of a Gaussian measure differs from the
list of properties of a von Neumann density operator only by one
condition: $\rm{Tr} \; \rho =1,$ for a density operator $\rho.$
Thus, for any covariance operator $B,$ its scaling $B/\rm{Tr B}$
can be considered as a density operator.

By using (\ref{CO}) we can easily find the Gaussian integral of
the quadratic form $f_A(\phi)$ defined by (\ref{QF}):
$$
\int_W f_A(\phi) dp_B(\phi)=\int_W (A\phi, \phi) dp_B (\phi)
$$
$$
=\sum^\infty_{i, j=1}(Ae_i, e_j) \int_W (e_i, \phi) (e_j, \phi) d
p_B(\psi)= \sum_{i, j=1}^\infty (A e_i, e_j)(Be_i, e_j),
$$
where $\{e_i\}$ is some orthonormal basis in $W.$ Thus
$\int_W f_A(\phi) dp_B (\phi)={\rm Tr}\; BA.$

\subsection{Complex case}

Let $Q$ and $P$ be two copies of a real Hilbert space. Let us
consider their Cartesian product $H=Q \times P,$  ``phase space,''
endowed with the symplectic operator $J= \left( \begin{array}{ll}
 0&1\\
-1&0
\end{array}
 \right ).$
Consider the class of Gaussian measures (with zero mean value)
which are invariant with respect to the action of the operator
$J;$ denote this class $S(H).$ It is easy to show that $p \in
S(H)$ if and only if its covariance operator commutes with the
symplectic operator, \cite{IZV}.

As always, we consider complexification of $H$ (which will be
denoted by the same symbol), $H=Q\oplus i P.$ The complex scalar
product is denoted by the symbol $\langle \cdot, \cdot \rangle.$
We  will also use the operation of complex congugation $*$ in complex Hilbert space $H,$ for  
$\phi= \phi_1+i \phi_2, \phi_1 \in Q, \phi_2 \in P,$ we set $
*(\phi) =\overline{\phi}= \phi_1 - i \phi_2.$  We will use the following trivial fact:
\begin{equation}
\label{LL}
\langle \overline{u}, \overline{v}\rangle =\langle v, u \rangle.
\end{equation} 

We introduce the complex covariance operator of a measure $p$ on
the complex Hilbert space $H$
$$
\langle Dy_1, y_2 \rangle = \int_H \langle y_1, \phi \rangle \langle \phi, y_2 \rangle d p (\phi).
$$
We also consider the complex Fourier transform of $p$
\begin{equation}
\label{Fourier} \tilde p (y) = \int_H exp \{i (\langle y, \phi
\rangle + \langle \phi, y \rangle )\} d p (\phi).
\end{equation}
Any $J$-invariant Gaussian measure on $H$ is determined by its
complex Fourier transform,\cite{IZV}:
$
\tilde{p} (y) = exp\{ - \langle Dy, y \rangle \}.
$

We remark that $J$-invariance is a strong constraint on the class
of Gaussian measures under consideration. Consider a measure $p$
on the Cartesian  product $H=Q\times P.$ Its real covariance
operator has the block structure\footnote{Little bit later we will
use the block structure of the complex covariance operator for a
measure defined on the Cartesian product of two {\it complex
Hilbert spaces.} The reader should be careful and not mix these
two totally different block structures!}

$ B= \left(
\begin{array}{ll}
B_{11} \; B_{12}\\
B_{21} \; B_{22}
 \end{array}
 \right ),
$
where $B_{11}^*= B_{11}, B_{22}^* = B_{22}, B_{12}^* =B_{21}.$
Consider also its complex covariance operator $D.$ It can be
realized as acting in the Cartesian product of two real Hilbert
spaces and in such a representation it also has the block
structure $D_{\rm{real}}= \left(
\begin{array}{ll}
\;L \;\;C\\
-C \;\; L
 \end{array}
 \right ).
$ It was shown in \cite{IZV} that $L=B_{11}+B_{22}$ and
$C=B_{12}-B_{21}.$ It also was shown that if measure is
symplectically invariant then $B_{11}=B_{22}, B_{21}=- B_{12}.$
Thus in the latter case the complex and real covariance operators
are coupled in a simple way: $D_{\rm{real}}=2B.$

\medskip

{\bf Lemma 1.} For any measure $p \in S(H)$ the following representation takes place
\begin{equation}
\label{K2}
\int_H \langle  \xi_1, \phi\rangle \langle \eta_1, \phi\rangle \langle \phi, \xi_2 \rangle \langle \phi, \eta_2 \rangle dp (\phi) = \langle D \xi_1, \eta_2\rangle \langle D \eta_1, \xi_2 \rangle + \langle D \xi_1, \xi_2 \rangle \langle D \eta_1, \eta_2 \rangle
\end{equation}

To prove this formula, one should differentiate  the Fourier
transform (\ref{Fourier}) four times.

\medskip

Let $H_1$ and $H_2$ be two complex Hilbert spaces and let
$D_{21} \in {\cal L} (H_1, H_2), D_{12} \in {\cal L} (H_2, H_1).$
Then $D_{21} \otimes D_{12} \in {\cal L} (H_1 \otimes H_2, H_2
\otimes H_1).$ Let us consider the permutation operator $\sigma:
H_2 \otimes H_1 \to H_1 \otimes H_2, \sigma (\phi_2 \otimes
\phi_1) = \phi_1 \otimes \phi_2.$ We remark that $\sigma \in {\cal
L} (H_2 \otimes H_1, H_1 \otimes H_2).$

Let $p$ be a measure on the Cartesian product $H_1 \times H_2$ of
two Hilbert spaces. Then its covariance operator has the block
structure 
\begin{equation}
\label{BLOCK}
D = \left( \begin{array}{ll}
 D_{11} & D_{12}\\
D_{21} & D_{22}\\
 \end{array}
 \right ),
 \end{equation}
where $D_{ii} : H_i \to H_i$ and $D_{ij}: H_j \to H_i.$ The
operator is self-adjoint. Hence $D_{ii}^* = D_{ii},$ and $D_{12}^*
= D_{21}.$

Let $H$ be a complex Hilbert space and let $\widehat A \in {\cal
L} (H, H).$ We consider its quadratic form (which will play an important role 
in our further considerations)
$$
\phi \to f_A(\phi) = \langle \widehat{A} \phi,\phi\rangle. 
$$
We make a trivial, but ideologically important remark:  $f_A: H \to H ,$ is a ``usual 
 function'' which is defined point wise. 

In the same way as in the real case we prove the equality
\begin{equation}
\label{QIY} \int_H f_A(\phi) dp_D (\phi)={\rm Tr}\; DA
\end{equation}

\medskip

{\bf Theorem 1.} {\it Let $p \in S(H_1 \times H_2)$ with the
(complex) covariance operator $D$ and let $\widehat A_i \in {\cal
L} (H_i, H_i), i= 1,2.$ Then}
\begin{equation}
\label{BL}
\int_{H_1 \times H_2} 
f_{A_1}( \phi_1) f_{A_2}(\phi_2) d p (\phi) = 
{\rm Tr} D_{11} \widehat A_1 \; {\rm Tr}D_{22} \widehat A_2 + 
{\rm Tr}D_{12} \widehat A_2 D_{21} \widehat A_1
\end{equation}

\medskip

This theorem is a consequence of the following general result:

{\bf Lemma 2.} {\it Let $p \in S(H)$ with the
(complex) covariance operator $D$ and let $\widehat A_i \in {\cal L} (H, H), i= 1,2.$ Then}
\begin{equation}
\label{YY1} \int_{H} f_{A_1}( \phi) f_{A_2}(\phi)   dp (\phi), = {\rm Tr} D
\widehat A_1 {\rm Tr} D \widehat A_2 + {\rm Tr} D \widehat A_2 D
\widehat A_1.
\end{equation}

{\bf Proof.} By Lemma 1 the integral can be represented as $$I =
\sum_{i_1 j_1} \sum_{i_2 j_2} \langle \widehat A_1 e_{i_1}, e_{j_1} \rangle \langle \widehat A_2 e_{i_2}, e_{j_2}\rangle$$
$$\times [\langle D e_{j_1}, e_{i_2} \rangle \langle D e_{j_2}, e_{i_1} \rangle + \langle D e_{j_1}, e_{i_1} \rangle \langle D e_{j_2}, e_{i_2} \rangle ] = I_1 + I_2,
$$
where $\{e_i\}$ is an orthonormal basis in $H.$
Here
$$
I_1= \sum_{i_1 i_2} \sum_{j_1} \langle \widehat A_1 e_{i_1},  e_{j_1} \rangle \langle e_{j_1}, D e_{i_2} \rangle \sum_{j_2} \langle \widehat A_2 e_{i_2}, e_{j_2} \rangle
\langle e_{j_2}, D e_{i_1} \rangle
$$
$$
 = \sum_{i_1 i_2} \langle D \widehat A_1  e_{i_1}, e_{i_2}\rangle \langle e_{i_2}, \widehat A_2^* D e_{i_1} \rangle$$
$$ =
\sum_i \langle D \widehat A_2 D \widehat A_1 e_{i_1}, e_{i_1} \rangle ={\rm Tr} D \widehat A_2 D \widehat A_1.
$$
$$
I_2 = \sum_{i_1 i_2} \sum_{j_2} \langle \widehat A_1 e_{i_1} e_{j_1} \rangle \langle e_{j_1}, D e_{i_1}\rangle \sum_{j_2} \langle \widehat A_2 e_{i_2} e_{j_2}\rangle \langle e_{j_2}, D e_{i_2}\rangle
$$
$$=
\sum_{i_1 i_2} \langle \widehat A_1 e_{i_1}, D e_{i_1}\rangle \langle \widehat A_2 e_{i_2},D e_{i_2}\rangle = {\rm Tr} D \widehat A_1 {\rm Tr} D_{A_2}.
$$

{\bf Proposition 1.} {\it Let conditions of Theorem 1 hold. Then}
\begin{equation}
\label{ZZ}
\int_{H_1 \times H_2} 
f_{A_1}( \phi_1) f_{A_2}(\phi_2)  
dp (\phi) = {\rm Tr} (D_{11} \otimes D_{22} + \sigma (D_{21} \otimes D_{12})) \widehat A_1 \otimes \widehat A_2.
\end{equation}

{\bf Proof.} It is sufficient to prove that $$
{\rm Tr} \sigma (D_{21} \otimes D_{12}) \widehat A_1 \otimes \widehat A_2
 = {\rm Tr} D_{12} \widehat A_2 D_{21} \widehat A_1.
$$
We have
$$
{\rm Tr} \sigma (D_{21} \otimes D_{12}) \widehat A_1 \otimes A_2 = \sum_{ij} \langle \sigma (D_{21} \otimes D_{12}) \widehat A_1 \otimes \widehat A_2 e_i \otimes f_j, e_i \otimes f_j \rangle $$
$$
=\sum_{ij} \langle D_{12} \widehat A_2 f_j  \otimes D_{21} \widehat A_1  e_i, e_i \otimes f_j \rangle =
\sum_{ij} \langle D_{12} \widehat A_2 f_j, e_i \rangle \langle D_{21} \widehat A_1 e_i, f_j \rangle.
$$

On the other hand, ${\rm Tr D_{12} \widehat A_2 D_{21} \widehat A_1}$
$$
= \sum_i \langle D_{12} \widehat A_2 D_{21} \widehat A_1 e_i, e_i \rangle
\sum_i \langle D_{21} \widehat A_1 e_i, A_2^* D_{21} e_i \rangle
\sum_{ij} \langle D_{21} \widehat A_1 e_i, f_j \rangle \langle f_j, A_2^* D_{21} e_i \rangle.
$$

\section{Vectors, operators, traces}

\subsection{Vector and operator realizations of the tensor product}

In quantum theory a pure state of a composite system is represented by a normalized
{\it vector} belonging to the tensor product $H_1\otimes H_2.$ On the 
 other hand, in functional analysis
it is common to use elements of $H_1\otimes H_2$ as operators.
The standard construction provides the realization of any vector $\Psi \in H_1\otimes H_2$
by a linear operator from $H_2^* \to H_1,$ where $H_2^*$ is the space dual to $H_2.$ For a vector
$Z= u\otimes v,$ one puts $\widehat{Z} x^* = x^*(v)u$ for $x^*\in H_2^*.$ This correspondence 
$Z\to \widehat{Z}$ is extended to isomorphism of $H_1\otimes H_2$ with the space of Hilbert-Smidt operators 
${\cal HS}(H_2^*, H_1)$ or (equivalently) with the space of anti-linear $HS$-operators from $H_2$ to $H_1.$
The main compication in coming considerations is related to our need to represent vectors from the tesor product by {\it linear} operators       
 from $H_2$ to $H_1.$ We present a new construction which seems to be unknown, so it may be interesting even 
from the purely mathematical viewpoint. We remind the definition of the $HS$-norm corresponding to the
 trace scalar product in ${\cal HS}(H_2, H_1).$ Take an arbitrary orthonormal basis $\{f_k\}$ in $H_2.$ Then  
 $$
 \langle \widehat{L_1}, \widehat{L_2} \rangle= \sum_k \langle \widehat{L_1}f_k, \widehat{L_2} f_k \rangle=
\rm{Tr}   \widehat{L_2}^* \widehat{L_1}, \; \Vert \widehat{L}\Vert_2^2 = \rm{Tr} \widehat{L}^* \widehat{L}
$$ 
(index 2 is typically used for the Hilbert-Schmidt norm).

\medskip

Let $\{e_j\}$ and $\{f_j\}$ be two orthonormal bases
in  $H_1$ and $H_2,$ respectively. Then $$\Psi = \sum_{ij}
\psi_{ij} e_i \otimes f_j, \psi_{ij} \in {\bf C},
$$ and $||\Psi||^2
= \sum_{ij} |\psi_{ij}|^2.$ We remark that, for an orthonormal  
basis, say  $\{f_j\},$ the system of complex conjugate vectors
$\{\overline{f}_j\}$ is also an orthnormal  
basis. Really, by (\ref{LL}) we get $\langle \overline{f}_j, \overline{f}_i \rangle=  
\langle f_i, f_j \rangle = \delta_{ij}.$ 

We set, for $\phi \in H_2,$
\begin{equation}
\label{AO}
\widehat \Psi \phi = \sum_{ij} \psi_{ij} \langle\phi_2, \overline{f}_j \rangle e_i.
\end{equation}
We emphasize that, for the vector $\Psi,$ the expansion with respect to the basis $\{e_i\otimes f_j\}$ was used.
In cotrast to this, in the expansion of the operator $\widehat \Psi$ the basis $\{e_i\otimes \bar{f}_j\}$ was used.
If the basis is real than the definition is essentially simplified:
\begin{equation}
\label{AOLK}
\widehat \Psi \phi = \sum_{ij} \psi_{ij} \langle\phi_2, f_j \rangle e_i.
\end{equation}

Take now a factorizable vector $\Psi= u\otimes v.$ It defines the rank one operator 
$$
\widehat \Psi \phi= \langle \phi, \bar{v}\rangle u= \langle v, \bar{\phi}\rangle u.
$$
To play with the definition, take $\Psi= c u\otimes v= (cu) \otimes v= u \otimes (cv),$
 where $c \in {\bf C}.$ On the one hand,  we have $\widehat \Psi \phi= c \langle \phi, \bar{v}\rangle u;$
 on the other hand, $\widehat \Psi \phi= \langle \phi, \overline{c v}\rangle u= \langle \phi, \bar{c} \bar{v}\rangle u=
c \langle \phi, \bar{v}\rangle u.$ Thus our definition is consistent with scaling by a complex constant.
Consider now very special, but at the same time very important case: $H_i=L_2({\bf R}^{n_i}), i=1,2.$
Here our definition gives the following representation (a special case of (\ref{AOLK})):
\begin{equation}
\label{AOLK1}
\widehat \Psi \phi(x) = \int \Psi(x,y) \phi(y) d y
\end{equation}
We proceed formally and use the real basis $\{e_x\otimes f_y\},$ where $e_x(t)= \delta(t-x)$ and 
$f_y(s)=\delta(s-y).$ Finally, we remind a result of theory of integral operators. The operator given by  (\ref{AOLK1})
is of the $HS$-type in $L_2$-spaces if and only if its kernel $\Psi(x,y)$ is square integrable. Thus the condition
$
\Psi \in L_2({\bf R}^{n_1}\times {\bf R}^{n_2})  
$
for a function $\Psi$ is equivalent to the $HS$-condition for the operator $\widehat{\Psi}.$ In the abstract form this fact will
be formulated in coming lemma.  

\medskip

{\bf Lemma 3.} {\it Each vector $\Psi \in H_1\otimes H_2$
determines (uniquely) operator $\widehat \Psi \in {\cal L} (H_2,
H_1)$ and, moreover, $\widehat \Psi \in {\cal HS}(H_1,H_2)$ and $||\widehat \Psi ||_2 =||  \Psi||$
(the norm of a vector coicides with the Hilbert-Schmidt norm of the corresponding operator)}

{\bf Proof.} We start with proof of corectness of definition (\ref{AO}), i.e., it does not depend on the choice of orthonormal bases.
Let $\{e_j^\prime\}$ and $\{f_j^\prime \}$ be two orthonormal bases in  $H_1$ and $H_2,$ respectively, which are in general different from
bases $\{e_j\}$ and $\{f_j\}.$  Consider the expansion  $\Psi= \sum_{ij}
\psi_{ij}^\prime e_i^\prime \otimes f_j^\prime.$ We remark that 
$\overline{f_j ^\prime}= \sum_k \langle \overline{f_j ^\prime}, \overline{f_k}\rangle \overline{f_k}=
\sum_k \langle f_k, f_j^\prime \rangle \overline{f_k},$ which can also be obtained by complex conjugation from the expansion
 $f_j^\prime= \sum_k \langle f_j^\prime, f_k\rangle f_k.$ 
In this basis the vector $\Psi$ defines the operator
\begin{equation}
\label{AO1}
\widehat{\Psi}^\prime \phi = \sum_{ij} \psi_{ij} ^\prime \langle \phi, \overline{f_j^\prime} \rangle e_i^\prime.
\end{equation}
We have
$$  
\widehat{\Psi}^\prime \phi= \sum_{ij} \sum_{nm} \psi_{nm}  \langle e_n, e_i^\prime \rangle
 \langle f_m, f_j ^\prime \rangle  \langle \phi, \sum_k \langle \overline{f_j^\prime} , \overline{f_k} \rangle \overline{f_k} \rangle
\sum_p \langle e_i^\prime, e_p \rangle e_p
$$
$$
 =  \sum_{nm} \psi_{nm} \sum_{kp}  \langle \phi, \overline{f_k} \rangle 
 \sum_i  \langle e_n, e_i^\prime \rangle 
 \langle e_i^\prime, e_p \rangle 
\sum_j \langle  \overline{f_k}, 
\overline{f_j^\prime} \rangle
\langle f_m, f_j^\prime \rangle  e_p.
$$
We remark 
$$
\sum_j \langle  \overline{f_k}, 
\overline{f_j^\prime} \rangle
\langle f_m, f_j^\prime\rangle=\sum_j \langle f_m, f_j^\prime\rangle \langle f_j^\prime, f_k\rangle= \delta_{km}
$$
Thus $
\widehat{\Psi}^\prime \phi= \sum_{nm} \psi_{nm}  \langle \phi, \overline{f_m} \rangle e_n.
$ 
It coincides with (\ref{AO}). We now apply the Cauchy-Bunyakovsky inequality to the expression 
(\ref{AO}) and obtain:
$
\Vert \widehat \Psi \phi\Vert^2 = \sum_{i} \vert \sum_{j}\psi_{ij} \langle\phi, \overline{f}_j \rangle\vert^2\leq
\sum_{i} \sum_{j} \vert \psi_{ij}\vert ^2  \sum_{j}\vert \langle\phi, \overline{f}_j \rangle\vert^2=
\Vert \Psi \Vert^2 \Vert \phi \Vert^2.
$
Thus this operator belongs to the space ${\cal L}(H_2, H_1).$ 
Finally, we show that it belongs even to the space ${\cal HS}(H_2, H_1):$
$
\Vert \widehat{\Psi} \Vert_2^2= \rm{Tr} \widehat{\Psi}^* \widehat{\Psi}=
\sum_k \langle  \widehat{\Psi} \bar{f}_k, \widehat{\Psi} \bar{f}_k\rangle=
 \sum_k \langle \sum_{i_1} \psi_{i_1 k} e_{i_1},  \sum_{i_1} \psi_{i_2 k} e_{i_2} \rangle=
  \sum_{ki}  \psi_{i k} \overline{\psi_{i k}}.$

\medskip

Moreover, it is easy to show that any operator $\widehat{L} \in {\cal HS}(H_2, H_1)$ can be represented as 
(\ref{AO}) for some vector $\Psi \in H_1\otimes H_2.$  In the combination with Lemma 3 this remark  implies:

\medskip

{\bf Corollary 1.} { The equality (\ref{AO}) establishes the isomorphism of Hilbert spaces $H_1\otimes H_2$ and 
${\cal HS} (H_2, H_1).$  

We now find the adjoint operator. We have:
$$
\langle \widehat{\Psi} y, x \rangle=
 \sum_{nm} \psi_{nm} \langle  y, \overline{f_m} \rangle \langle e_n, x \rangle=
\langle y, \sum_{nm} \overline{\psi_{nm}} \langle x, e_n \rangle \overline{f_m} \rangle.
$$
Thus 
$$
\widehat{\Psi}^*x=  \sum_{nm}\overline{\psi_{nm}} \langle x, e_n\rangle \overline{f_m}.
$$

\subsection{Operation of the complex conjugation in the space of self-adjoint operators}

Let $\widehat{A}\in {\cal L}_s(W),$ where $W$ is Hilbert space. We define 
``complex conjugate operator'' $\widehat{\bar{A}}$ by its bilinear form:
\begin{equation}
\label{AOC}
\langle   \widehat{\bar{A}} u, v \rangle = \langle \bar{v}, \widehat{A} \bar{u}\rangle.
\end{equation}
Let $\{f_j\}$ be an orthnormal basis in $W.$ We find the matrix of the operator 
$\widehat{B}\equiv \widehat{\bar{A}}$ with respect to this basis:
$
b_{ij}= \langle   \widehat{\bar{A}} f_i, f_j \rangle = \langle \bar{f}_j, \widehat{A} \bar{f}_i\rangle.
$
In the special case of the real basis, i.e., $\bar{f}_j= f_j,$ we have:
$$
b_{ij}=\overline{a_{ij}}= a_{ji}.
$$
Thus its matrix is given by the transposition of the matrix of $\widehat{A}.$ 
We will use the fact that the operator $ \widehat{\bar{A}}$ is self-adjoint (we remind that $\widehat{A}$ is self-ajoint):
\begin{equation}
\label{AOC1}
\langle   \widehat{\bar{A}} u, v \rangle = \langle \widehat{A} \bar{v},  \bar{u}\rangle=
\overline{\langle   \bar{u}, \widehat{A} \bar{v} \rangle} =
\overline{\langle  \widehat{\bar{A}} v, u  \rangle} =
\langle  u,  \widehat{\bar{A}} v  \rangle.
\end{equation}
We will also use the fact that, for a positiv operator $\widehat N,$ the operator $\widehat{\bar{N}}$ is also positiv:
\begin{equation}
\label{AOC1Y}
\langle  \widehat{\bar{N}} u, u \rangle=\langle  \bar{u}, \widehat{N} \bar{u}\rangle=
\langle  \widehat{N} \bar{u}, \bar{u} \rangle.
\end{equation}
Consider the quadratic form of the complex conjugate operator $\widehat{\bar{A}}$ of a self-adjoint operator 
$\widehat{A}:$
\begin{equation}
\label{AOC1N}
f_{\bar{A}}(\phi) =\langle \widehat{\bar{A}} \phi, \phi\rangle= \langle  \bar{\phi},  \widehat{A}\bar{\phi}\rangle=
f_{A}(\bar{\phi}).
\end{equation}
Consider the group $\{e, *\},$ where $* $ is the operation of complex conjugation in a compex Hilbert space $W.$
It induces the action in the space of real valued functions on 
$W: f\to \bar{f},$ where $\bar{f}(\phi)= f(\bar{\phi}).$ We hope that the symbol $\bar{f}$ will not be misleading.
Only real valued functions are under consideration. Thus it cannot be mixed with the operation of complex conjugation
on the range of values.

Theorem 1 implies (for  $\widehat A_i \in {\cal
L}_s (H_i, H_i), i= 1,2):$ 
\begin{equation}
\label{BLZ}
\int_{H_1 \times H_2} 
f_{A_1}( \phi_1) \bar{f}_{A_2}(\phi_2) d p (\phi_1, \phi_2) = 
{\rm Tr} D_{11} \widehat A_1 \; {\rm Tr}D_{22} \widehat{\bar{A}_2} + 
{\rm Tr}D_{12} \widehat{\bar{A}_2} D_{21} \widehat A_1
\end{equation}

\subsection{The basic operator equality}

{\bf Lemma 4.} Let $\Psi\in H_1\otimes H_2.$  
Then, for any pair of operators $\widehat{A}_j \in {\cal L}_s (H_j), j= 1,2, $
\begin{equation}
\label{01}
\rm{Tr} \widehat{\Psi} \widehat{\bar{A}}_2 \widehat{\Psi}^* \widehat{A}_1=
\langle \widehat{A}_1 \otimes \widehat{A}_2 \rangle_\Psi \equiv
\langle \widehat{A}_1 \otimes \widehat{A}_2 \Psi, \Psi \rangle.
\end{equation}

{\bf Proof.} We have: 
$$
\rm{Tr} \widehat{\Psi} \widehat{\bar{A}}_2\widehat{\Psi}^*  \widehat{A}_1=
\sum_k \langle  \widehat{\Psi} \widehat{\bar{A}}_2\widehat{\Psi}^*  \widehat{A}_1 e_k, e_k \rangle=
\sum_k \sum_{ij} \psi_{ij}  \langle \widehat{\bar{A}}_2\widehat{\Psi}^*  \widehat{A}_1 e_k, \overline{f_j} \rangle \delta_{ik}
$$
$$
=\sum_{ij} \psi_{ij}  \langle \widehat{\bar{A}}_2\widehat{\Psi}^*  \widehat{A}_1 e_i, \overline{f_j} \rangle=
\sum_{i_1 j_1} \sum_{i_2 j_2} \psi_{i_1 j_1} \overline{\psi_{i_2j_2}}  \langle  \widehat{A}_1 e_{i_1}, e_{i_2} \rangle 
\langle \overline{f_{j_2}}, \widehat{\bar{A}}_2 \overline{f_{j_1}} \rangle.
$$
By (\ref{AOC1}) and (\ref{AOC}) we obtain 
$\langle \overline{f_{j_2}}, \widehat{\bar{A}}_2 \overline{f_{j_1}} \rangle=
\langle \widehat{\bar{A}}_2 \overline{f_{j_2}},  \overline{f_{j_1}} \rangle =
\langle f_{j_1}, \widehat{A}_2 f_{j_2} \rangle=
\langle \widehat{A}_2 f_{j_1},  f_{j_2} \rangle.$ Thus
$$
\rm{Tr} \widehat{\Psi} \widehat{\bar{A}}_2\widehat{\Psi}^*  \widehat{A}_1=
\sum_{i_1 j_1} \sum_{i_2 j_2} \psi_{i_1 j_1} \overline{\psi_{i_2j_2}}  \langle  \widehat{A}_1 e_{i_1}, e_{i_2} \rangle 
\langle \widehat{\bar{A}}_2 f_{j_1},  f_{j_2} \rangle.
$$
On the other hand,  we obtain:
$$
\langle \widehat{A}_1 \otimes \widehat{A}_2 \Psi, \Psi \rangle =
\sum_{i_1 j_1} \sum_{i_2 j_2}  \psi_{i_1 j_1} \overline{\psi_{i_2 j_2}}  
\langle \widehat{A}_1 e_{i_1},  e_{i_2} \rangle \langle
\widehat{A}_2 f_{j_1}, f_{j_2}\rangle=  \rm{Tr} \widehat{\Psi} 
\widehat{\bar{A}}_2\widehat{\Psi}^*  \widehat{A}_1.
$$

If $\Psi$ is normalized by 1, then the right-hand side of equality (\ref{01}) is nothing else than 
average of the observable  $\widehat{C}= \widehat{A}_1 \otimes \widehat{A}_2$ describing 
correlations between measurement of observables $\widehat{A}_1$ and  $\widehat{A}_2$ on subsystems 
$S_1$ and $S_2$ of a composite system $S=(S_1, S_2)$ which is prepared in the state $\Psi.$
On the other hand, the left-hand side of 
equality (\ref{01}) has the form of the second term in the right-hand side of  formula (\ref{BL}) 
 giving Guassian integral of the product of two quadratic forms corresponding to operators  
$\widehat{A}_1$ and  $\widehat{A}_2.$ These mathematical coincidences provide a possibility to couple
quantum correlations with classical Gaussian correlations, by selecting the covariance operator of the prequantum 
Gaussian distribution (corresponding to quantum state $\Psi)$  in the right way. 

\subsection{Operator representation of reduced density operators}

{\bf Lemma 5.} {\it For any  vector $\Psi \in H_1\otimes H_2,$ the following equality holds:}
\begin{equation}
\label{00}
{\rm Tr}_{H_2} \Psi \otimes \Psi = \widehat \Psi \widehat \Psi^*.
\end{equation}

{\bf Proof.} The operator $\widehat \Psi \widehat \Psi^*$ acts on vector $x\in H_1$ as 
$$
\widehat{\Psi} \widehat{\Psi}^* x= \sum_{n m} \overline{\psi_{nm}} \langle x, e_n\rangle  \widehat{\Psi} \overline{f_m}
=\sum_{n m} \sum_{kl} \psi_{kl} \overline{\psi_{nm}} \langle  \overline{f_m}, \overline{f_l}\rangle 
\langle x, e_n\rangle  e_k
$$
$$
=\sum_{n k} [\sum_{m} \psi_{km} \overline{\psi_{nm}} ] \langle x, e_n\rangle  e_k.
$$
Thus its bilinear form is given by
$$
\langle \widehat{\Psi} \widehat{\Psi}^* x, y \rangle= 
\sum_{n k} [\sum_{m} \psi_{km} \overline{\psi_{nm}} ] \langle x, e_n\rangle  \langle e_k, y\rangle.
$$

Take an orthogonormal basis $\{f_m\}$  in $H_2.$  
 The bilinear form of the operator ${\rm Tr}_{H_2} \Psi \otimes \Psi$ is given by 
$$
\langle {\rm Tr}_{H_2} \Psi \otimes \Psi x, y \rangle=
\sum_m \langle {\rm Tr}_{H_2} \Psi \otimes \Psi x \otimes f_m, y \otimes f_m \rangle 
$$
$$
=\sum_m \sum_{kl} \langle x, e_k \rangle   \langle e_l, y\rangle  \langle e_k\otimes f_m, \Psi \rangle 
 \langle\Psi, e_l \otimes f_m\rangle 
=\sum_{kl} [\sum_{m} \psi_{lm} \overline{\psi_{km}} ] \langle x, e_k \rangle  \langle e_l, y \rangle.
$$   
 By setting $k \to n, l \to k$ we obtain the coincidence of two bilinear forms and hence the operators.
 
 \medskip
 
If the vector $\Psi$ is normalized, then $\rho_\Psi= \Psi\otimes\Psi$ is the corresponding 
density operator and $\rho^{(1)}_\Psi$ is the $H_1$-reduced density operator. By   
(\ref{00}) we obtain
\begin{equation}
\label{00YU}
\rho^{(1)}_\Psi=  \widehat \Psi \widehat \Psi^*.
\end{equation}

Unfortunately,  for the $H_1$-reduced density operator $\rho^{(2)}_\Psi,$ similar statement is not true; in general,
\begin{equation}
\label{00YU1}
\rho^{(2)}_\Psi \not=  \widehat \Psi^* \widehat \Psi.
\end{equation}

Let find the bilinear form of the operator $\widehat \Psi^* \widehat \Psi: H_2 \to H_2:$  
  $$
  \widehat \Psi^* \widehat \Psi y= \sum_{nm} \psi_{nm} \langle y, \overline{f_m} \rangle  \widehat{\Psi}^* e_n=
 \sum_{nm, ij} \psi_{nm} \overline{\psi_{ij}} \langle y, \overline{f_m} \rangle \langle e_n, e_i \rangle \overline{f_j}
$$
$$
= \sum_{mj} [\sum_i \psi_{im} \overline{\psi_{ij}}] \langle y, \overline{f_m} \rangle \overline{f_j}.
$$
\begin{equation}
\label{00JKTR}
 \langle  \widehat{\Psi}^* \widehat \Psi y, u \rangle= 
\sum_{mj} [\sum_i \psi_{im} \overline{\psi_{ij}}] \langle y, \overline{f_m} \rangle 
\langle \overline{f_j}, u\rangle.
\end{equation}
However, for any orthonornal basis $\{e_m\}$ in $H_1,$ we obtain:
 $$\langle {\rm Tr}_{H_1} \Psi \otimes \Psi y, u \rangle
= \sum_m  \langle \Psi \otimes \Psi e_m \otimes y, e_m \otimes  u\rangle
$$
$$
=\sum_{mjl}  \langle e_m \otimes f_j, \Psi \rangle \langle \Psi, e_m \otimes f_l \rangle
\langle y, f_j \rangle \langle f_l, u \rangle
$$
\begin{equation}
\label{00JK}
\sum_{jl} [ \sum_m \psi_{ml} \overline{\psi_{mj}}] \langle y, f_j\rangle\langle f_l, u \rangle.
\end{equation}
By setting $m \to i, j \to m, l\to j$ in (\ref{00JK})  we obtain 
\begin{equation}
\label{00JK1} 
\langle {\rm Tr}_{H_1} \Psi \otimes \Psi y, u\rangle =
\sum_{mj} [ \sum_i \psi_{ij} \overline{\psi_{im}}] \langle y, f_m\rangle\langle f_j, u \rangle.
\end{equation}
Comparing (\ref{00JKTR}) and (\ref{00JK1}) we see that in general they do not coincide.
 To show this, let us take the real basis $\{f_m\}.$ In this case   (\ref{00JKTR}) has the form:
\begin{equation}
\label{00JKTRY}
 \langle  \widehat{\Psi}^* \widehat \Psi y, u \rangle = 
\sum_{mj} [\sum_i \psi_{im} \overline{\psi_{ij}}] \langle y, f_m \rangle 
\langle f_j, u\rangle.
\end{equation}
Thus in this special basis the matrix elements of operators  
$\widehat{\Psi}^* \widehat \Psi$ and  ${\rm Tr}_{H_2} \Psi \otimes \Psi$ are coupled via the complex 
conjugation. In general, we have: 

\medskip

{\bf Lemma 5$^*.$} {\it For any  vector $\Psi \in H_1\otimes H_2,$ the following equality holds for the operator 
 $\widehat{T}={\rm Tr}_{H_1} \Psi \otimes \Psi$:}
\begin{equation}
\label{00PQ}
 \widehat{\bar{T}} = \widehat \Psi^* \widehat \Psi.
\end{equation}

{\bf Proof.}  We now present proof which is not based on matrix elements. In particular, it illustrates well 
features of complex conjugate operators. For $\widehat{T}$ defined in the formulation, we obtain:
$$
\langle \widehat{\bar{T}} y, u\rangle= \langle \bar{u}, \widehat{T} \bar{y} \rangle=
\overline{\langle \widehat{T} \bar{y}, \bar{u} \rangle}.
$$
By (\ref{00JK}) we get:
$$
\langle \widehat{\bar{T}} y, u\rangle= \sum_{mj}[\sum_i \overline{\psi_{ij}} \psi_{im}] \langle f_m, \bar{y}\rangle \langle \bar{u},
f_j\rangle.
$$
By (\ref{LL}) $\langle\bar{u}, v\rangle = \langle \bar{v}, u \rangle$ and hence  
$\langle\bar{u}, f_j \rangle = \langle \bar{f_j}, u \rangle$ and $\langle u,\bar{v}\rangle= 
\langle v,\bar{u}\rangle$ and hence $\langle f_m, \bar{y}\rangle = \langle y, \bar{ f_m}\rangle.$
Thus 
$$
\langle \widehat{\bar{T}} y, u\rangle= \sum_{mj}[\sum_i \overline{\psi_{ij}} \psi_{im}] \langle y, \bar{ f_m}\rangle 
\langle \bar{f_j}, u \rangle =    \langle  \widehat{\Psi}^* \widehat \Psi y, u \rangle.
$$
In particular, if $\Psi$ is normalized (pure quantum state) then we obtain 
\begin{equation}
\label{00PT}
 \overline{\rho_\Psi^{(2)}} = \widehat \Psi^* \widehat \Psi.
\end{equation}

\medskip

{\bf Lemma 6.} {\it Let $\rho$ be a density operator in a Hilbert space $W.$ Then, for any 
$\widehat{A}\in {\cal L}_s(W),$ the following equality holds:}
\begin{equation}
\label{00PT1}
\rm{Tr} \; \overline{\rho}\widehat{\bar{A}}= \rm{Tr} \; \rho \widehat{A}.
\end{equation}

{\bf Proof.} Let $\{e_k\}$ be an orthonormal basis in $W.$ Then 
$$ 
 \rm{Tr} \; \overline{\rho}\widehat{\bar{A}}=\sum _k \langle\overline{\rho}\widehat{\bar{A}} e_k, e_k\rangle=
 \sum _k \langle  \bar{e}_k, \rho \overline{\widehat{\bar{A}}  e_k}\rangle
 $$
 $$
 =\sum _k \langle  \rho \bar{e}_k,  \overline{\widehat{\bar{A}}  e_k}\rangle=
 \sum _k \langle   \widehat{\bar{A}}  e_k,  \overline{\rho \bar{e}_k}\rangle=
 $$
 $$
 = \sum _k \langle \rho \bar{e}_k, \widehat{A}  \bar{e}_k\rangle=
 = \sum _k \langle \widehat{A}  \rho \bar{e}_k, \bar{e}_k\rangle.
$$
Since the trace does not depend on the choice of a basis, we can select the real basis.
Hence, it was proved that    
$$
 \rm{Tr} \; \overline{\rho}\widehat{\bar{A}}=\rm{Tr} \; \widehat{A}  \rho.
$$
Finally, we will prove that $$ \rm{Tr} \; \widehat{A}  \rho= \rm{Tr} \;   \rho \widehat{A}.$$
Take the basis consisting of the eigenvectors of the density operator:  $\rho=\sum_k p_k e_k\otimes  e_k.$
Then
$$
\rm{Tr} \;   \rho \widehat{A}= \sum_k \langle \rho \widehat{A} e_k, e_k\rangle= \sum_k p_k 
\langle \widehat{A} e_k, e_k\rangle
=\sum_k  \langle \widehat{A} \rho e_k, e_k\rangle.
$$

{\bf Corollary 2.} {\it  Let $\Psi\in H_1\otimes H_2$ be normalized (pure quantum state). Then}
\begin{equation}
\label{00P}
\rm{Tr} \;  \widehat \Psi^* \widehat \Psi \widehat{\bar{A}}=
\rm{Tr}  \; \rho_\Psi^{(2)} \widehat{A}. 
\end{equation}

\section{Classical random field description}
\label{LLKK}

\subsection{Ensemble of noncomposite quantum systems}

In what-follows  random vectors taking values in a Hilbert space are called {\it random fields.}
This definition is motivated by consideration of the Hilbert space $H=L_2({\bf R}^m)$ 
of square integrable
functions.

Let $\phi (\omega)$  denote a Gaussian random
field  in a complex Hilbert space $H.$    
Everywhere below we consider Gaussian
random fields with probability distributions of the class $S(H).$ 
 The covariance operator of a random field  is defined as 
the covariance operator of its probability distribution.

The correspondence between QM and PCSFT in the case of a single quantum system with 
the state space $H$ is established in the following way:

1). Density operators (statistical states of QM) are identified with covariance operators of 
prequantum random fields, $\rho \mapsto D.$

2). Self-adjoint operators (quantum observables) are identified with quadratic functionals,
$\widehat{A} \mapsto f_A.$

The equality (\ref{QIY}) can be written as 
$$
Ef_A(\phi(\omega))= 
\rm{Tr} \rho \widehat{A} \equiv \langle \widehat{A} \rangle_\rho. 
$$
It establishes the correspondence between PCSFT-averages and QM-averages.
This story was presented in \cite{EKHR}- \cite{KH6}. Now we modify it. Originally 
the source of coming modification was purely mathematical -- 
to solve the problem of positive definiteness in theory for composite systems, see
section \ref{TTHHJJ}. However, 
it happens that a natural physical interpretation can be provided. 

To escape measure-theoretic difficulties, at the moment we proceed  in the 
finite-dimensional Hilbert space. We will come back to the real physical case
(for infinite dimension) in section \ref{TTHHJJ}, see Proposition 4.
Let $\rho$ be a density operator.  
Set $D= \rho + \alpha I,$ where $I$
is the unit operator and $\alpha>0.$ Consider the  Gaussian random
vector $\phi(\omega)$ with the covariance operator $D.$
The additional term $\alpha I$ we can consider as ($\alpha$-scaling of) the Gaussian 
normal distribution.  It describes {\it spatial white noise} when the dimension of the space goes 
to infinity.

We have $ E f_A(\phi(\omega))= 
\rm{Tr} \rho
\widehat{A} + \alpha \rm{Tr} \widehat{A}. $ In this model
(modification of PCSFT created in \cite{KH1}-- \cite{KH6})
 quantum average can be obtained as a shift of classical average:
\begin{equation}
\label{CO3}
\langle \widehat{A} \rangle_\rho=
E f_A(\phi(\omega)) - \alpha \rm{Tr} \widehat{A}.
\end{equation}
The shift is generated by the presence of the {\it background Gaussian noise}
(say ``zero point field'', cf. SED, \cite{R2}-\cite{BOY}).  Thus QM-average can be 
considered as simply normalization  of average with respect to a prequantum random field.
Normalization consists of substraction of the contribution of the background field.
While in the finite-dimensional case the use of such a normalization is just a matter of test,
in the infinite-dimensional case it becames very important. If quantum observable is represented
by an operator $\widehat{A}$ which is not of the trace class, then the normalization 
$\rm{Tr} \widehat{A}= \infty.$ In other words the quadratic form $f_A(\phi)$ is not 
integrable, cf. Proposition 4, with respect to the probability distribution $p_D,$ where   
 $\rho + \alpha I.$ Of course, it is a pure theoretical problem. In real experimental practice
 we are able to measure only observables represented by operators of finite ranks, see von 
Neumann \cite{VN}. Other observables (in particular, all observables given by operators 
with continuous spectra) are just mathematical idealizations. Nevertheless, it is convenient 
to have a theory which is able to operate with such quantities as well. From the PCSFT-viewpoint 
QM is such a theory. Thus in our approach QM has some analogy with QFT, but all 
divergences are regularized from the very beginning by choosing a special representation of 
classical averages.  

\subsection{Ensemble of composite quantum systems}

Consider a composite quantum system $S=(S_1, S_2).$  Here $S_j$ has the state space
$H_j,$ a complex Hilbert space. 
Let $\phi_1 (\omega)$ and $\phi_2 (\omega)$ be two Gaussian random
fields,   in Hilbert spaces $H_1$ and $H_2,$ respectively. Consider
the Cartesian product of these Hilbert spaces,
$H_1 \times H_2,$ and the vector Gaussian
random field   $\phi (\omega) = (\phi_1 (\omega), \phi_2 (\omega)) \in H_1 \times H_2.$
In the case under consideration its covariance operator 
 has the block structure given by (\ref{BLOCK}). Set 
$$
\langle f_{A_1}, f_{A_2}\rangle= = E f_{A_1}  \bar{f}_{A_2}= 
\int_{H_1\times H_2} f_{A_1}(\phi_1)  \bar{f}_{A_2} (\phi_2) dp (\phi_1, \phi_2).
$$
Set also
$$
{\rm cov} \; (f_{A_1}, f_{A_2})= \langle f_{A_1}, f_{A_2}\rangle - \langle f_{A_1}\rangle \langle \bar{f}_{A_2}\rangle
$$
Equalities (\ref{BLZ}) and (\ref{01})  imply 

\medskip

{\bf Proposition 2.} {\it Let $\widehat{A}_i \in {\cal L}_s(H_i), i=1,2$ 
and let $\Psi \in H_1\otimes H_2$ with the unit norm. Then, for any Gaussian random field  
$\phi(\omega)$ in $H_1\times H_2$ with the covariance matrix $D$   such that the 
non-diagonal block 
\begin{equation}
 \label{HHUURR}
D_{12}= \widehat{\Psi}
\end{equation} 
the following equality takes place:}
 \begin{equation}
 \label{00T5}
{\rm cov} \; (f_{A_1}, f_{A_2}) = (\widehat{A}_1 \otimes \widehat{A}_2 \Psi, \Psi) \equiv
\langle \widehat{A}_1 \otimes \widehat{A}_2 \rangle_\Psi.
\end{equation}

This equality establishes coupling between quantum and classical correlations.
In the next section we will unify classical descriptions for a single system, section
\ref{LLKK},  and a composite system.

\subsection{Making  consistent  PCSFT-models for ensembles of noncomposite and composite 
systems}
\label{TTHHJJ}

Operators $D_{ii}$ are responsible for averages of
functionals  depending only on one of components
of the vector random field $\phi (\omega).$ In particular,
$E f_{A_1} (\phi_1) (\omega)) = {\rm Tr} D_{11} \widehat{A}_1$
and $E \bar{f}_{A_2} (\phi_2) (\omega)) = {\rm Tr} D_{22} \widehat{\bar{A}}_2.$
We will  construct such a random field that these ``marginal averages'' 
will match those given
by QM. For the latter, we have:
$$
\langle \widehat{A}_1 \rangle_\Psi = (\widehat{A}_1 \otimes I_2 \Psi, \Psi) =
{\rm Tr} \rho_\Psi^{(1)} \widehat{A}_1,
\langle \widehat{A}_2 \rangle_\Psi = 
(I_1 \otimes \widehat{A}_2 \Psi, \Psi)={\rm Tr} \rho_\Psi^{(2)} \widehat{A}_2,
$$
where $I_i$ denotes the unit operator in $H_i, i=1,2.$
By equality (\ref{00YU}) the first average can written as
$$
\langle \widehat{A}_1 \rangle_\Psi={\rm Tr} (\widehat{\Psi} \widehat{\Psi}^*) \widehat{A}_1.
$$
By equality (\ref{00PT1}) the second average can be written as 
$$
\langle \widehat{A}_2 \rangle_\Psi = {\rm Tr} \overline{\rho_\Psi^{(2)}} \widehat{\bar{A}}_2
$$
and, finally, by (\ref{00PT})
$$
\langle \widehat{A}_2 \rangle_\Psi
  = {\rm Tr} (\widehat{\Psi}^* \widehat{\Psi}) \widehat{\bar{A}}_2,
$$
 Thus it
would be natural to take
 \[D = \left( \begin{array}{ll}
 \widehat{\Psi} \widehat{\Psi}^* & \; \; \widehat{\Psi} \\
 \; \; \widehat{\Psi}^* & \widehat{\Psi}^*\widehat{\Psi}
 \end{array}
 \right ).
 \]
Its off-diagonal block reproduces correct quantum correlations between systems $S_1$ and $S_2$ and its 
diagonal blocks produce  correct quantum averages for system $S_1$ and system $S_2.$

However, in general (i.e., for an arbitrary pure state $\Psi)$
this operator is not positively defined. Therefore (in general) it could not be chosen as the
covariance operator of a random field. Let us consider a modification  which will be positively defined
and such that quantum and classical averages will be coupled by a simple rule.
Thus from quantum averages one can easily find classical averages and vice versa.

\medskip

{\bf Proposition 3.} {\it For any normalized vector $\Psi \in H_1\otimes H_2,$
 the operator}
\begin{equation}
\label{YYYO}
D = \left( \begin{array}{ll}
 (\widehat{\Psi} \widehat{\Psi}^* + I_1/4) & \; \; \; \; \; \; \; \widehat{\Psi}\\
\; \;  \; \; \; \; \; \widehat{\Psi}^* & (\widehat{\Psi}^*\widehat{\Psi} +
I_2/4)
 \end{array}
 \right )
\end{equation}
 {\it is positively defined.}

{\bf Proof.}  For any vector $\phi=(\phi_1, \phi_2) \in \Psi \in H_1\times H_2,$ we have:
$ (\tilde{D}_\Psi \phi, \phi) = ||\widehat{\Psi}^*
\phi_1 ||^2 + \frac{||\phi_1 ||^2}{4} + (\widehat{\Psi} \phi_2, \phi_1) +
(\widehat{\Psi}^* \phi_1, \phi_2) + ||\widehat{\Psi} \phi_2 ||^2 + \frac{||\phi_2 ||^2}{4} \geq (|| \Psi \phi_1 ||^2 - || \Psi^* \phi_1 || || \phi_2 || + \frac{|| \phi_2
||^2}{4}) + (|| \widehat{\Psi} \phi_2 ||^2 - ||\phi_1 || ||\widehat{\Psi} \phi_2|| +
\frac{||\phi_1||^2}{4}) \geq 0.$ Thus operator $D$ is positively defined.\footnote{Of course,
the same effect can be approached by adding $\alpha I$  for $\alpha \geq 1/4.$}

\medskip

We continue to proceed in the finite-dimensional case 
(to escape the problem of existence of $\sigma$-additive Gaussian
measure on infinite-dimensional space).

For the  Gaussian measure with covariance operator (\ref{YYYO}), we have:
$
\langle f_{A_1}\rangle= \langle \widehat{A}_1 \rangle_\Psi - \rm{Tr}\widehat{A}_1/4, 
\langle \bar{f}_{A_2}\rangle= \langle \widehat{A}_2 \rangle_\Psi - \rm{Tr}\widehat{A}_2/4.
$ 
 These relations for averages together with relation for correlations (\ref{00T5})
provide coupling between PCSFT and QM.

In the infinite-dimensional case Gaussian distribution with he covariance
operator given by (\ref{YYYO}) is not $\sigma$-additive. 

To make it $\sigma$-additive
one should consider a rigged Hilbert space: ${\bf H}_+ \subset {\bf H} \subset 
{\bf H}_-,$  where $H= H_1\times H_2,$ and  both embedding operators
are of the Hilbert-Schmidt class.

\medskip

{\bf Proposition 4.} {\it For any normalized vector $\Psi \in  H_1\otimes H_2,$
 the operator(\ref{YYYO}) determines the $\sigma$-additive Gaussian distribution on
 ${\bf H}_-$ or equivalently the random field $\phi(\omega)$ valued in ${\bf H}_-.$ For 
 trace class
operators $\widehat{A}_i: H_i \to H_i, i=1,2,$ 
equalities (\ref{00T5}) and (\ref{YY1}) take place.}

\medskip

To prove this proposition, one should repeat the previous proofs, 
existence of traces is based on the trace class condition for 
 of operators $A_i$  (and not the trace class feature of the 
 covariance  operator of a Gaussian measure). The crucial difference with the finite
dimensional case is that the prequantum random field takes values not in the Cartesian 
product  $H=H_1\otimes H_2,$ but in its Hilbert-Schmidt extension. 
For mathematical details, I would like
 to recommend the excellent short book of A. V. Skorohod \cite{[59]}, see
 also \cite{[60]}-\cite{[63]} for applications to mathematical physics.

  \section{Classical (Hilbert valued) stochastic process corresponding to 
 Schr\"odinger's evolution}

We again start our considerations by considering the finite-dimensional case.
Since we do not try to go beyond QM, but only reproduce its
predictions, we use Schr\"odinger's equation for dynamics of the
``wave function''\footnote{So, we do not try to modify this
equation, cf. \cite{KH3}}. We only change the interpretation of the
$\Psi$-function of the composite system. Thus we start with
Schr\"odinger's equation for a composite system $S=(S_1,S_2):$
\begin{equation}
\label{SCHR} i\frac{d \Psi}{d t} (t)= \widehat{H} \Psi(t), \; \Psi(0)=
\Psi_0,
\end{equation}
where $\widehat{H}$ is Hamiltonian of $S.$

Hence, at the instant
$t,$ the covariance matrix of the prequantum random field (vector
in the finite-dimensional case) $\phi(t, \omega)$ has the form:
\begin{equation}
\label{SJ} 
D(t)= \left( \begin{array}{ll}
 (\widehat{\Psi(t)} \widehat{\Psi(t)}^* + I_1/4) & \; \; \; \; \; \; \; \widehat{\Psi(t)}\\
\; \;  \; \; \; \; \; \widehat{\Psi(t)}^* &
(\widehat{\Psi(t)}^* \widehat{\Psi(t)} + I_2/4)
 \end{array}
 \right )
\end{equation}
The following fundamental question (having both mathematical and
physical counterparts) immediately arises:

{\it ``Can one construct a stochastic process (valued in the Cartesian
product $H_1 \times H_2)$ such that at each $t\in [0,
\infty)$ its covariance matrix coincides with $D_(t)?"$}

\subsection{Bernoulli type process}

The formal mathematical answer is yes! It is easy to construct
such a stochastic process. Take space  $\Omega= \prod_{t \in [0,
\infty)} H_1 \times H_2$ as the space of random parameters, points of
this space $\omega=(\omega_t)$ can be considered as functions
$\omega: [0, \infty) \to H_1 \times H_2,$ trajectories. Consider the
family of Gaussian measures $p_t$ on $H_1 \times H_2$ having zero mean
value and covariance operators $D_(t), t \in [0, \infty).$ Consider now (on
$\Omega)$ the direct product of these measures, $P= \prod_{t \in
[0, \infty)} p_t.$  

\medskip

{\bf Proposition 5.} {\it  Let $\phi(t, \omega)$ be a stochastic process
having the probability distribution $P$ on $\Omega.$  
Then, for any pair of vectors $y_1, y_2 \in H_1 \times H_2$ and any
instant of time $t \geq 0,$ 
\begin{equation}
\label{SJ1}
E \langle y_1, \phi(t, \omega)\rangle  \langle \phi(t,\omega), y_2
\rangle =\langle  D(t)  y_1, y_2 \rangle,
\end{equation}
where $D(t)$ is given by (\ref{SJ1}).}

\medskip

Existence of this stochastic process is a consequence of famous Kolmogorov's theorem.
The equality (\ref{SJ1}) is a consequence of the definition of probability $P$ on $\Omega.$

Thus there exists a prequantum classical stochastic process
inducing the Schr\"odinger evolution for any composite system
prepared initially in a pure state. One may say that, for a
composite system, Schr\"odinger's equation describes dynamics of
the nondiagonal block of the covariance matrix of such a
prequantum stochastic process.

This story becomes essentially more complicated after the remark
that  such a  {\it  prequantum process is not uniquely determined}
by  the $D(t)!$ To determine uniquely a Gaussian process
(up to natural equivalence), one should define not only covariance
for each instant of time, i.e, \\ $E \langle y_1, \phi(t,
\omega)\rangle  \langle \phi(t,\omega), y_2 \rangle,$ but so
called {\it covariance kernel} $D(t,s):$
$$
E \langle y_1, \phi(t, \omega)\rangle , \langle \phi(s,\omega),
y_2 \rangle=\langle  D(t,s)  y_1, y_2 \rangle.
$$
However, the formalism of QM does not provide such a possibility.
It is a consequence of the trivial fact (but of the great
importance, cf. von Neumann \cite{VN}) that Schr\"odinger's
equation for a composite system is dynamics with respect to a {\it
single time parameter $t,$} common for both subsystems, and not
with respect to a pair of time parameters $(t,s)$ corresponding to
internal times of subsystems.

Nevertheless, one may feel that the process existing due to Proposition 5 
is not adequate to the real physical situation.  Since its probability distribution 
$P$ is the direct product of probabilities corresponding to different instances of time, 
it is the {\it Bernoulli process.} Its value at the instance of time $t$ is totally independent 
from the previous behavior. Although this process provides right averages for each instance
of time, it is hard to believe that real physical dynamics of e.g. an electron is of the Bernoulli-type
(and for any Hamiltonian $\widehat{H}).$ We are looking for more realistic stochastic
processes.  

\subsection{Stochastic  (local) dynamics in the absence of interaction} 

We restrict our consideration to dynamics in the absence of
interactions between $S_1$ and $S_2$ after the preparation
procedure. Thus we are interested in propagation of initially
correlated random fields (vectors in the finite-dimensional case).
Although it is a rather special dynamics, it plays an important
role in quantum foundations. In particular, it describes the
evolution of entanglement in the EPR-Bohm type experiments. Thus
we consider Hamiltonian
\begin{equation}
\label{SJ2}
\widehat{H}=\widehat{H_1}\otimes I_2 + I_1\otimes \widehat{H_2},
\end{equation}
where $\widehat{H_j}$ is Hamiltonian of $S_j$ (here we use QM
terminology).

\medskip

{\bf Lemma 7.} {\it Let Hamiltonian have the form (\ref{SJ2}).
Then} 
\begin{equation}
\label{EQW} \widehat{\Psi(t)}=e^{-i \widehat{H_1}t}
\widehat{\Psi_0} e^{-i \widehat{H_2}t}.
\end{equation}
\begin{equation}
\label{EQW1} \widehat{\Psi(t)}^* =e^{i \widehat{H_2}t}
\widehat{\Psi_0}^* e^{-i \widehat{H_1}t}.
\end{equation}

{\bf Proof.} In this case $$\Psi(t)=e^{-it (\widehat{H_1}\otimes I_2 +
I_1\otimes \widehat{H_2})} \Psi_0.$$ 
We expand the initial state $\Psi_0:$ 
$
\Psi_0= \sum_{ij}^k \psi_{ij}  e_i\otimes f_j.
$
Then
$$
\Psi(t)=\sum_{ij} \psi_{ij}  e^{-i
\widehat{H_1}t} e_j\otimes e^{-i \widehat{H_2}t} f_j.
$$
Thus, for $v \in H_2,$ we get
$$
\widehat{\Psi(t)} v = \sum_{ij}^k \psi_{ij}  \langle v, e^{i
\widehat{H_2}t}\bar{f}_j \rangle  e^{-i \widehat{H_1}t} e_j =   e^{-i
\widehat{H_1}t} [\sum_{ij}^k \psi_{ij}  \langle e^{-i
\widehat{H_2}t} v, \bar{f}_j \rangle  e_j].
$$

\medskip

By using Lemma 7 we prove:

{\bf Lemma 8.} {\it Let the condition of Lemma 7 hold. Then}
\begin{equation}
\label{EQW2} \widehat{\Psi(t)} \widehat{\Psi(t)}^* =e^{-i
\widehat{H_1}t}  \widehat{\Psi_0} \widehat{\Psi_0}^* e^{i
\widehat{H_1}t}.
\end{equation}
\begin{equation}
\label{EQW3} \widehat{\Psi(t)}^* \widehat{\Psi(t)}= e^{i
\widehat{H_2}t}  \widehat{\Psi_0}^* \widehat{\Psi_0} e^{-i
\widehat{H_2}t} .
\end{equation}

\medskip

Finally, we obtain:

{\bf Lemma 9.} {\it   Let the condition of Lemma 7 hold. 
Then the operator $D(t)$ given by (\ref{SJ}) can be represented in the form:}
\begin{equation}
\label{EQW2Z}
D(t)= \left( \begin{array}{ll}
 (e^{-i\widehat{H_1}t} \widehat{\Psi_0} \widehat{\Psi_0}^* e^{i \widehat{H_1}t} +I/4)
& \; \; \; \; \; \; \; e^{-i \widehat{H_1}t}
\widehat{\Psi_0}  e^{-i \widehat{H_2}t}\\
\; \;  \; \; \; \; \; e^{ i\widehat{H_2}t} \widehat{\Psi_0}^*
e^{i \widehat{H_1}t} &(e^{i\widehat{H_2}t} \widehat{\Psi_0}^*
\widehat{\Psi_0} e^{-i \widehat{H_2}t} +I/4)
 \end{array}
 \right )
 \end{equation}

By using this representation it is easy to prove:

\medskip

{\bf  Proposition 6.} {\it   Let the condition of Lemma 7 hold. 
Then the operator $D(t)$ given by (\ref{SJ})  is the covariance operator 
(for each instance of time $t)$ of the vector process with coordinates
\begin{equation}
\label{EQWX} \phi_1(t, \omega)= e^{-i
\widehat{H_1}t}\xi_{01}(\omega), \phi_2(t, \omega)= e^{i
\widehat{H_2}t} \xi_{02} (\omega),
\end{equation}
where the initial random vector $\xi_0(\omega)=(\xi_{01} (\omega),
\xi_{02} (\omega))$ is Gaussian with zero mean value  and the
covariance operator$D(0).$} 

{\bf Proof.}  We will find not only the covariance operator for a
fixed instant of time, but even the covariance kernel. We have, for
any pair of vectors $u, w \in H_1,$
$$E \langle u, \phi_1(t, \omega)\rangle  \langle
\phi_1(s,\omega), w \rangle= E \langle e^{i
\widehat{H_1}t}u, \xi_{01}(\omega) \rangle
\langle\xi_{01}(\omega)), e^{i \widehat{H_1}s} w \rangle
$$
$$
=\langle e^{-i \widehat{H_k}s} (\widehat{\Psi_0}^*
\widehat{\Psi_0} +I_1/4)e^{i \widehat{H_1}t} u, w
\rangle.
$$
The same calculations can be done for the second diagonal block.
Thus diagonal blocks of the covariance operator of the stochastic
process given by (\ref{EQWX}) coincide with diagonal blocks of the
operator $D(t).$ We now consider nondiagonal blocks. Let
now $u \in H_1, v \in H_2.$ We have:
$$
E \langle u, \phi_1(t, \omega)\rangle  \langle \phi_2(s,\omega), v
\rangle= E \langle e^{i \widehat{H_1}t}u, \xi_{01}(\omega) \rangle
\langle\xi_{02}(\omega)), e^{-i \widehat{H_2}s} v \rangle
$$
$$
= \langle\widehat{\Psi_0}^* e^{i \widehat{H_1}t} u,  e^{-i
\widehat{H_2}s}  v \rangle= \langle e^{i \widehat{H_2}s}
\widehat{\Psi_0}^* e^{i \widehat{H_1}t} u, v \rangle.
$$
The same calculations cane be done for the second nondiagonal
block. Thus the covariance kernel has the form $D(t,s)=$
\begin{equation}
\label{EQWXj}
 \left( \begin{array}{ll}
 e^{-i\widehat{H_1}s} \widehat{\Psi_0} \widehat{\Psi_0}^* e^{i \widehat{H_1}t} +e^{i \widehat{H_1}(t-s)}I/4
& \; \; \; \; \; \; \; e^{-i \widehat{H_1}s}
\widehat{\Psi_0}  e^{i \widehat{-H_2}t}\\
\; \;  \; \; \; \; \; e^{i \widehat{H_2}s} \widehat{\Psi_0}^*
e^{i \widehat{H_1}t} &e^{i\widehat{H_2}s} \widehat{\Psi_0}^*
\widehat{\Psi_0} e^{-i \widehat{H_2}t} +e^{-i \widehat{H_2}(t-s)}
I/4
 \end{array}
 \right )
 \end{equation}
And hence, for $t=s, D(t,t)=D(t).$

\medskip

We emphasize that dynamics (\ref{EQWX}) by
itself is purely deterministic, stochasticity is generated by
initial conditions. One might say that this process describes
propagation of uncertainty of preparation.

 We remark that  Propositions 5 and 6 provide two different 
stochastic processes. The covariance kernel (\ref{EQWXj})
 differs from the covariance kernel of the
 process which has been constructed by considering the product of
 Gaussian distributions $p_t.$ The latter has the covariance
 kernel 
$$
D(t, s)=D(t) \delta(t-s).
$$ 
For this process, its  realization at different instants of time are independent.The
 process defined by (\ref{EQWX}) contains nontrivial dependence
 between its realizations at different times. I think that it is
 closer to the real physical situation.

 The following interesting problem arises:

 \medskip

 {\it To construct a stochastic process for an arbitrary
 Hamiltonian, such that in the case of the
 absence of interactions this construction gives
 the process (\ref{EQWX}).}

 At the moment I am not able to solve this problem.

\subsection{Stochastic nonlocal dynamics}

We now consider another classical stochastic process reproducing 
dynamics of quantum correlations. 

{\bf Proposition 7.} {\it Let operator $D(t)$ be defined by (\ref{SJ}). 
Then the stochastic process 
\begin{equation}
\label{PR}
\xi (t, \omega) = \sqrt{D(t)} \eta_0 (\omega),
\end{equation}
where $\eta_0 (\omega) \in H_1 \times H_2$ is distributed $N (0, I),$ 
has the covariance operator $D (t)$ for any $t \geq 0.$}

{\bf Proof.} Let $y_1, y_2 \in H_1 \times H_2.$ Then 
$$
E \langle y_1, \xi (t, \omega)\rangle \langle \xi (t, \omega), y_2 \rangle = 
E \langle \sqrt{D (t)} y_1, \eta_0 (\omega) \rangle 
\langle \eta_0 (\omega),$$ $$ \sqrt{D (t)} y_2 \rangle = 
\langle \sqrt{D(t)} y_1, \sqrt{D (t)} y_2\rangle 
= \langle D(t) y_1, y_2 \rangle.
$$
It is clear that the covariance kernel is given by 
\begin{equation}
\label{PR1}
D (t, s) = \sqrt{D(s) D (t)}.
\end{equation}

We remind that, for  Hamiltonian without interaction, we constructed the 
stochastic process $\phi(t, \omega)$ given by 
(\ref{EQWX}). In general stochastic processes, $\xi (t, \omega)$ and 
$\phi(t, \omega)$ given by 
(\ref{PR}) and (\ref{EQWX}) do not coincide:

 We can write the process (\ref{EQWX}) as $\phi(t) = V (t) \sqrt{D (0)} \eta_0,$ 
where $V(t) = {\rm diag} (e^{-it\widehat H_1},e^{it \widehat H_2}).$ Hence, $$D_\phi (t, s) = V(s) D (0) V (t)^*.$$
On the other hand, the covariance kernel of process (\ref{PR}) is given by 
$
D_\xi (t,s) = \sqrt{D(s) D (t)}.
$ 
We remark that 
$
D(t) = V(t) D (0) V (t)^*.
$ 
Hence, 
$
\sqrt{D(t)} = V(t) \sqrt{D(0)}V(t)^*.
$ 
Thus
$$
D_\xi (t,s)= V(s) \sqrt{D(0)} V^* (s) V (t) \sqrt{D(0)} V^*(t).
$$
We remark that $V^* (s) V (t)  \ne I, t\not=s.$

The process $\xi (t, \omega)$ is nonlocal in the following sense. Its component 
$\xi_1 (t, \omega)$ is guided not only by the Hamiltonian of $S_1,$  but also of $S_2;$
the same is valid for $\xi_2 (t, \omega).$ Thus PCSFT (at least at the moment) cannot provide 
a definite answer on locality of the prequantum world. Quantum correlations can be produced by local as well 
as nonlocal prequantum stochastic processes.

\medskip

{\bf Proposition 8.}
{\it In the case of  Hamiltonian  without interaction, see (\ref{SJ2}),
the stochastic process (\ref{PR}) can be represented in the form:
\begin{equation}
\label{PRR1}
\xi_1 (t) = e^{-it\hat H_1} Q_{11}^0 e^{it \hat H_1} \eta_{01} + 
e^{-it \hat H_1} Q_{12}^0 e^{-it \hat H_2} \eta_{02},
\end{equation}
\begin{equation}
\label{PRR2}
\xi_2 (t) = e^{it \hat H_2} Q_{21}^0 e^{it\hat H_1} \eta_{01} + 
e^{it \hat H_2} Q_{22}^0 e^{-it \hat H_2} \eta_{02},
\end{equation}
where
 \[ \sqrt{D(0)}= \left( \begin{array}{ll}
Q_{11}^0 & Q_{12}^0\\
Q_{21}^0 & Q_{22}^0\\
 \end{array}
 \right ),
 \] 
 and $\eta_0 (\omega) \in N (0, I).$}

{\bf Proof.} For example, take $y_1, y_2 \in H_1$ and consider average 

$$E \langle y_1, \xi (t, \omega) \rangle \langle \xi (s, \omega), y_2 \rangle
$$ 
$$
=E \langle e^{-it \hat H_1} Q_{11}^0 e^{it \hat H_1} y_1, \eta_{02}(\omega)\rangle 
\langle \eta_{01} (\omega), e^{-is \hat H_1} Q_{11}^0 e^{is \hat H_1} y_2 \rangle 
$$
$$
+ E \langle e^{it \hat H_2} (Q_{12}^0)^* e^{it \hat H_1} y_1, \eta_{02}(\omega)\rangle
\langle \eta_{02} (\omega), e^{is \hat H_2} (Q_{12}^0)^* e^{is \hat H_1} y_2\rangle
$$
$$
+ E \langle e^{-it \hat H_1} Q_{11}^0 e^{it \hat H_1} y_1, \eta_{10}(\omega)\rangle 
\langle \eta_{20} (\omega), e^{is \hat H_2} (Q_{12}^0)^* e^{is \hat H_1} y_2\rangle
$$
$$
+ E \langle e^{it \hat H_2} (Q_{12}^0)^* e^{it \hat H_1} y_1, \eta_{10}(\omega)\rangle
 \langle \eta_{20} (\omega), e^{-is \hat H_1} Q_{11}^0 e^{is \hat H_1} y_2\rangle.
$$
Two last terms are equal to zero, since $$E \langle z_1, \eta_{10} (\omega) \rangle \langle \eta_{20} (\omega), z_2 \rangle =0$$ for any pair $z_1 \in H_1, z_2 \in H_2.$ The first two give us
$$
\langle e^{-it \hat H_1} Q_{11}^0 e^{it \hat H_1} y_1, 
e^{-is \hat H_1} Q_{11}^0 e^{is \hat H_1} y_2 \rangle
$$
$$
+\langle e^{it \hat H_2} (Q_{12}^0)^* e^{it \hat H_1} y_1, 
e^{is \hat H_2} (Q_{12}^0)^* e^{is \hat H_1} y_2 \rangle.
$$
Thus 
$$
D_{11}(t, s) = e^{-is \hat H_1} Q_{11}^0 e^{i(s-t) \hat H_1}
 Q_{11}^0 e^{it \hat H_1}+ e^{-is \hat H_1} 
Q_{12}^0 e^{i(t-s) \hat H_2} (Q_{12}^0)^* e^{it \hat H_1}.
$$
\medskip
Representation (\ref{PRR1}), (\ref{PRR2}) implies that even in the absence of interaction between 
the subsystems $S_1$ and $S_2$ of the system $S$ the dynamics of $S_1$ depends on the 
Hamiltonian $\widehat H_2$ and vice versa. It can be interpreted as a sign of "action at the 
distance". Thus the stochastic process $\xi(t)$ can be considered as "nonlocal" --
opposite to the process $\phi(t)$  given by (\ref{EQWX}). 

\medskip

\subsection{Infinite-dimensional case}

To proceed in the infinite-dimensional case, one should consider  a rigged 
Hilbert space: ${\bf H}_+ \subset {\bf H} \subset  {\bf H}_-,$  
where $H= H_1\times H_2,$ and  both embedding operators
are of the Hilbert-Schmidt class. Stochastic processes take values in the Hilbert space $H_-$ 
(and not $H=H_1 \times H_2).$  All previous results are valid for any unitary dynamics
$\Psi(t)= U(t) \Psi_0.$

This paper was written under support of the grant ``Mathematical Modeling'' 
of V\"axj\"o university and the grant QBIC of Tokyo University of Science.
It was presented at ``Feynman Festival'', June, 2009; the author 
would like to thank Vladimir Manko for his critical comments which improved understanding 
of the model.


\begin{thebibliography}{99}

\bibitem{AKK} Khrennikov, A.(ed): Foundations of Probability and Physics.
Series PQ-QP: Quantum Probability and White Noise Analysis 13.
WSP, Singapore (2001)

\bibitem{KHC2}  Khrennikov, A.(ed): Quantum Theory:
Reconsideration of Foundations. Ser. Math. Model.  2, V\"axj\"o
University Press, V\"axj\"o  (2002); electronic volume:
http://www.vxu.se/msi/forskn/publications.html

\bibitem{ADC2} Adenier, G.,  Khrennikov, A. and Nieuwenhuizen, Th.M. (eds.):
Quantum Theory: Reconsideration of Foundations-3. American
Institute of Physics, Ser. Conference Proceedings 810, Melville,
NY (2006)

\bibitem{AKK5} Adenier, G., Fuchs, C.
and Khrennikov, A.(eds): Foundations of Probability and Physics-3.
American Institute of Physics, Ser. Conference Proceedings 889,
Melville, NY (2007)


\bibitem{AKK6} Accardi, L.,   G. Adenier, C.A. Fuchs, G. Jaeger, A. Yu. Khrennikov,
J.-A. Larsson, S. Stenholm (eds.): Foundations of Probability and
Physics-5, American Institute of Physics, Ser. Conference
Proceedings,  1101, Melville, NY (2009)


 \bibitem{KKK} Hess, K., Michielsen, K., De Raedt, H.: 
Possible experience: From Boole to Bell.  {\it EPL},   {\bf 87},  60007 (2009).


\bibitem{KKK1} Hess, K.: Modeling experiments using quantum and Kolmogorov probability.
{\it J. of Physics: Condensed Matter}, {\bf 20}, 454207  (2009).


\bibitem{IK}  A. Yu. Khrennikov, {\it Interpretations of Probability.} De
Gruyter, Berlin (2009),  second edition (completed).


\bibitem{K} Kolmogoroff, A. N.:  Grundbegriffe der Wahrscheinlichkeitsrechnung. Springer Verlag, Berlin (1933);
English translation: Kolmogorov, A.N.: Foundations of the
Probability Theory. Chelsea Publishing Company, New York (1956)

\bibitem{VN} J. von Neumann, {\it Mathematical Foundations of Quantum
Mechanics,} Princeton Univ. Press, Princeton, N.J., 1955.

\bibitem{GR} P. Busch, M. Grabowski, P. Lahti, {\it Operational Quantum
Physics,} Springer Verlag,Berlin, 1995.

\bibitem{Ozawa} M. Ozawa,  Conditional probability and a posteriori states in quantum mechanics,
{\it Publ. Res. Inst. Math. Sci.} {\bf  21},  279-295 (1985).

\bibitem{CK} A. Yu. Khrennikov, {\it Contextual approach to quantum
formalism,} Springer, Berlin-Heidelberg-New York, 2009.

\bibitem{EKHR} A. Yu. Khrennikov, Einstein's dream. Proceedings of Conference
{\it The nature of light: What are photons?} C. Roychoudhuri, A.
F. Kracklauer, K. Creath. Proceedings of SPIE, {\bf 6664}, 2007,
666409-1 -- 666409-9.

\bibitem{KH1}  A. Yu. Khrennikov,   ``Prequantum classical statistical model with
infinite dimensional phase-space,'' {\it J. Phys. A: Math. Gen.}
{\bf 38}, pp. 9051-9073, 2005.

\bibitem{KH2}  A. Yu. Khrennikov,   ``Generalizations of quantum mechanics
induced by classical statistical field theory,'' {\it Found. Phys.
Letters} {\bf 18}, pp. 637-650, 2005.

\bibitem{KH3}  A. Yu. Khrennikov, ``Nonlinear Schr\"odinger equations from prequantum
classical statistical field theory,'' {\it Physics Letters A} {\bf
357}, pp. 171-176, 2006.

\bibitem{KH4}  A. Yu. Khrennikov, ``Prequantum classical statistical field theory: Complex
representation, Hamilton-Schr\"odinger equation, and
interpretation of stationary states,'' {\it Found. Phys. Lett.}
{\bf 19}, pp. 299-319, 2006.

\bibitem{KH5}  A. Yu. Khrennikov, ``On the problem of hidden variables for quantum field theory,''
{\it Nuovo Cimento} B {\bf 121}, pp.  505-515, 2006.

\bibitem{KH6} A.  Yu. Khrennikov, Born's rule from classical random fields, {\it
Physics Letters} A,  {\bf 372},  N 44, 6588-6592 (2008).

\bibitem{R2}  De la Pena, L. and  Cetto, A. M.:  The quantum dice: An
introduction to stochastic electrodynamics.  Kluwer, Dordrecht
(1996)

\bibitem{B9} A. Casado, T. Marshall, E. Santos, {\it J. Opt. Soc. Am.} B
{\bf 14}, pp.  494-205, 1997.

\bibitem{10} G. Brida, M. Genovese, M. Gramegna, C. Novero and E. Predazzi,
{\it Phys. Lett} A {\bf 299}, pp.  121-141, 2002.

\bibitem{NH} Nieuwenhuizen, Th. M.:   Classical phase space density for
relativistic electron. In: Adenier, G.,  Khrennikov, A. and Nieuwenhuizen, Th.M. (eds.)
Quantum Theory: Reconsideration of Foundations-3. American Institute of Physics, Ser.
Conference Proceedings, vol. 810,  pp. 198-210. Melville, NY (2006)



\bibitem{BOY} Boyer, T. H.: A brief survey of stochastic electrodynamics.
In: Barut, A. O. (ed)  Foundations of Radiation Theory and Quantum
Electrodynamics, pp. 141-162.  Plenum, New York (1980)


\bibitem{C1} Scully, M. O.  and  Zubairy, M. S.: Quantum optics.
 Cambridge University Press,  Cambridge (1997)

\bibitem{C2} Louisell,  H. H.:  Quantum statistical properties of radiation.
 J. Wiley, New York (1973)

\bibitem{C3} Mandel,  L.  and Wolf,  E.: Optical
coherence and quantum optics.   Cambridge University Press,
Cambridge  (1995)

\bibitem{N} Nelson, E:  Quantum fluctuation
Princeton Univ. Press,  Princeton (1985)

\bibitem{MD} Davidson, M.:  J. Math. Phys. 20,  1865-1870 (1979)

\bibitem{MD1} Davidson,  M.:  Stochastic models of quantum
mechanics - a perspective. In: Adenier, G., Fuchs,   C.  and
Khrennikov, A.  (eds.)   Foundations of Probability and Physics-4.
American Institute of Physics, Ser. Conference Proceedings, vol.
889, pp. 106--119. Melville, NY (2007)

\bibitem{H1} `t Hooft, G.:   Quantum mechanics and determinism.
hep-th/0105105 (2001)

\bibitem{H2} `t Hooft, G.:   Determinism beneath quantum
mechanics. quant-ph/0212095 (2002)

\bibitem{EL} 't Hooft, G.: The free-will postulate in quantum
mechanics. quant-ph/0701097 (2007)

\bibitem{EL1} Elze,  T.: The attractor and the quantum states.
arXiv: 0806.3408 (2008)


\bibitem{Rus} Rusov, V. D. , Vlasenko, D. S., and  Mavrodiev, S.Cht.: 
Quantization in classical mechanics and reality of Bohm's psi-field.
arXiv:0906.1723 (2009)

\bibitem{Kis} Kisil, V.:  A quantum-classical brackets from p-mechanics.
Europhys. Lett. 72 (6), 873-?879 (2005)


\bibitem{M1}  V.  I.  Manko,  {\it J. of Russian Laser Research}, 
{\bf 17},  579-584 (1996).

\bibitem{Manko2}   V. I. Manko and E. V. Shchukin, 
{\it J. Russian Laser Research}, {\bf 22}, 545-560 (2001).

\bibitem{Manko1} M. A. Manko, V. I. Manko, R. V. Mendes,  
{\it J. Russian Laser Research}, {\bf 27}, 507-532.

\bibitem{M2}  S.  De Nicola,  R.  Fedele, M. A. Man'ko and V. I. Man'ko,
Quantum tomography, wave packets, and solitons. 	{\it J. of Russian Laser Research,}
{\bf 25}, 1071-2836,  2004.

\bibitem{M3}  O.  V. Manko and V. I. Manko,  {\it J. Russian Laser Research}, {\bf 25}, 477-492 (2004).

\bibitem{EPL} A.  Khrennikov, Entanglement's dynamics from classical stochastic process.
{\it Europhysics Letters}, {\bf 88}, 40005.1-6 (2009). 

\bibitem{IZV} A.  Khrennikov, To quantum averages through asymptotic expansion
of classical averages on infinite-dimensional space. {\it J. Math.
Phys.},  {\bf 48} (1), Art. No. 013512 (2007).

\bibitem{[59]} A. V. Skorohod, Integration in Hilbert space. Springer-Verlag,
Berlin, 1974. 

\bibitem{[60]} A. L. Daletski and S. V. Fomin, Measures and
differential equations in infinite-dimensional spaces. Kluwer,
Dordrecht, 1991.

\bibitem{[61]}  S. Albeverio, and M. R\"ockner, {\it Prob. Theory and Related
Fields} {\bf 89}, 347 (1991).

\bibitem{[63]} S. Albeverio, R. Hoegh-Krohn, {\it Phys. Lett.} {\bf B 177},
175 (1989).

\end{thebibliography}
\end{document}